\documentclass[prl,article,twocolumn,preprintnumbers,superscriptaddress]{revtex4}  

\usepackage{enumerate}
\usepackage{amssymb}
\usepackage{amsmath}
\usepackage{feynmf}
\usepackage{slashed}
\usepackage{natbib}
\usepackage{graphicx}
\usepackage[pdftex,bookmarks,linktocpage,pdfpagelabels,plainpages=false,hyperfigures,linkcolor=blue,citecolor=blue]{hyperref} 
\hypersetup{colorlinks=true}

\usepackage{mathrsfs,amssymb}  
\usepackage{cancel}
\usepackage[normalem]{ulem}

\newcommand\be{\begin{equation}}
\newcommand\ee{\end{equation}}

\newcommand\bea{\begin{eqnarray}}
\newcommand\eea{\end{eqnarray}}

\usepackage{graphicx}
\usepackage{xcolor}
\usepackage{amsmath,amssymb,amsthm,bm,multirow,longtable}

\renewcommand{\epsilon}{\varepsilon}
\renewcommand{\phi}{\varphi}

\newcommand{\Mwh}{M_{W\!H}}
\newcommand{\Mbh}{M_{BH}}

\theoremstyle{plain}

\theoremstyle{definition}
\allowdisplaybreaks

\newcommand{\baas}{{Bull. of the AAS}}  

\begin{document}

\bibliographystyle{apsrev4-1}

\title{
The Sound of Clearing the Throat: Gravitational Waves from a Black Hole Orbiting in a Wormhole Geometry}
\author{James B. Dent }
 \email{ jbdent@shsu.edu}
\affiliation{Department of Physics, Sam Houston State University, Huntsville, TX 77341, USA}
\author{William E. Gabella  }
\email{ b.gabella@vanderbilt.edu  }
 \author{ \\ Kelly Holley-Bockelmann   }
 \email{ k.holley@vanderbilt.edu}
\author{Thomas W. Kephart}
\email{tom.kephart@gmail.com}
\affiliation{Department of Physics and Astronomy, Vanderbilt University, Nashville, TN 37235, USA}
\date{\today}
\begin{abstract}
Current ground-based gravitational wave detectors are tuned to capture the collision of compact objects such as stellar origin black holes and neutron stars; over 20 such events have been published to date. Theoretically, however, more exotic compact objects may exist, collisions of which should also generate copious gravitational waves. In this paper, we model the inspiral of a stellar mass black hole into a stable, non-spinning, traversable wormhole, and find a characteristic waveform -- an {\it  anti-chirp} and/or {\it burst}  --  as the black hole emerges, i.e., outspirals, into our region of the Universe. This novel waveform signature may be useful in searches for wormholes in future gravitational wave data or used to constrain  possible  wormhole  geometries  in  our Universe. 
\end{abstract}
\maketitle


\section{Introduction}
The discovery of black hole-black hole (BH-BH) mergers \cite{Abbott:2016blz},  mergers of black holes with neutron stars (BH-NS) \cite{TheLIGOScientific:2017qsa}, and NS-NS mergers \cite{Abbott:2020uma} has opened the new branch of gravitational wave (GW) astronomy.   As well as testing strong-field general relativity, GWs have unveiled a class of stellar origin black holes in the $30 -60 \, M_{\odot}$ range\cite{PhysRevX.9.031040}, underlining the promise of GW observations to open new discovery space. 

Beyond inspiraling binary black hole and neutron star systems, it has been suggested that gravitational waves can be used to detect a variety of objects ranging from cosmic defects \cite{Vilenkin:2000jqa} to exoplanets \cite{Wong:2018amf}. The possibility of detecting exotic compact objects such as wormholes (WHs) \cite{Wheeler:1955zz,Wheeler:1962} from an inspiral has also been put forth, with studies implying that such a scenario could produce a signal that is observationally distinguishable from other inspiraling binary systems \cite{Cardoso:2016rao}.
 
Thus far, investigations of binary systems within a wormhole spacetime have focused on the quasi-normal modes produced during the ringdown phase of the inspiral and the subsequent echo that is produced in the horizonless geometry (see for example \cite{Cardoso:2016oxy,Bueno:2017hyj,Aneesh:2018hlp,Abedi:2018npz,Abedi:2020sgg,Kirillov:2020cre}). In this work we argue that, in addition to the initial ringdown phase and echo, a massive object inspiraling within a traversable wormhole geometry, and passing through the throat, could exhibit  novel features with an unmistakable observational signature. Namely, we will follow the orbit of an object as it proceeds through the wormhole, outspirals into another region of spacetime, and then (in the case of an object gravitationally bound to the wormhole in the spacetime region across the throat) back into ours. This will produce a gap within the post-chirp GW signal when the object has passed through the throat. Continuing to follow the orbital process, however, the object will eventually pass back into our spacetime region, leading to an ``anti-chirp" signal characterized by a decreasing frequency profile with time (this can be a short burst of a signal when the time for the object to begin its new infall phase is small enough that the WH does not execute a complete orbit), and subsequent repetitions of the chirp/anti-chirp cadence could result with diminishing amplitude from GW energy loss. Alternatively, in our region of spacetime we could observe an anti-chirp without a preceding chirp signal as a BH emerges through a WH throat. This could be followed by a chirp signal in the case of a bounded orbit, or remain an isolated anti-chirp for an unbounded situation. Our goal in the present work is to explore a simple traversable wormhole geometry that can lead to these types of signatures. 

\section{Wormhole Construction}

\begin{figure}[ht]
	\includegraphics[width=0.5\textwidth]{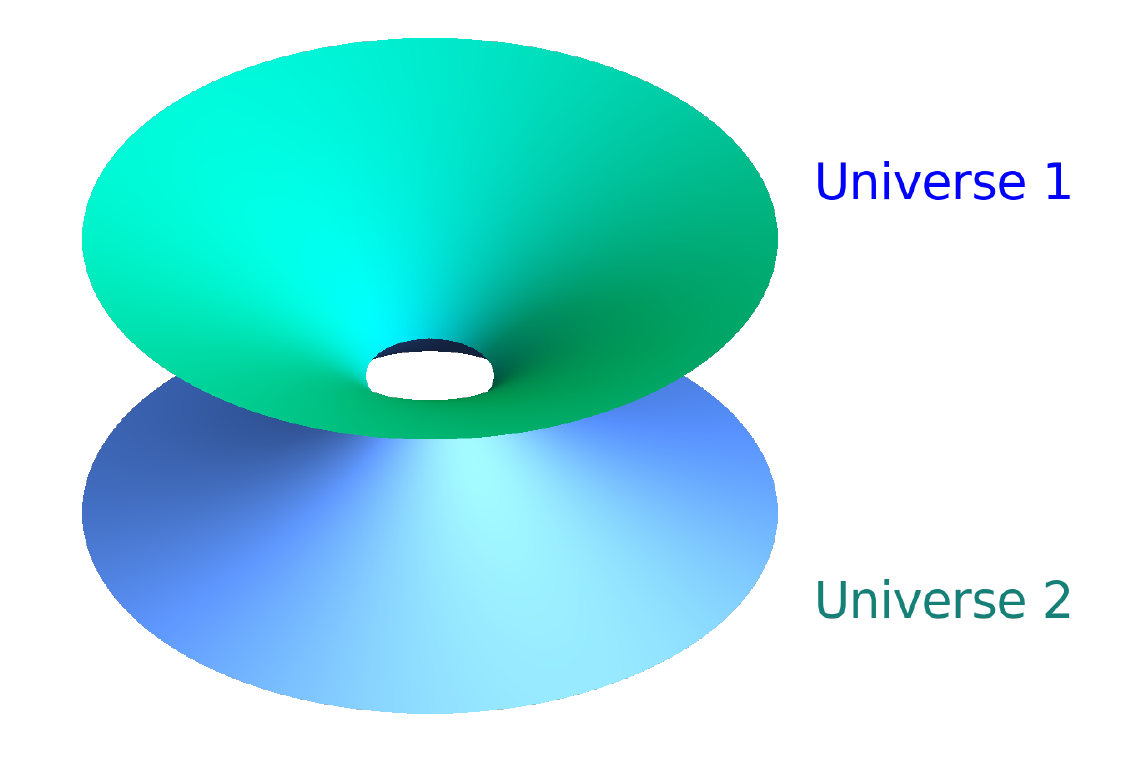}
	\caption{\footnotesize{The embedding diagram for a Schwarzschild-like wormhole, with the asymptotically flat spacetimes at the top and at the bottom of the figure, and a throat at radius $r=3 \Mwh$, where $\Mwh$ is the effective mass for the spacetime curvature on both sides of the throat.  There is no spacetime for $r<3 \Mwh$.}}
	\label{fig:WormholeEmbed}  
\end{figure}

Here we consider a type of Lorentzian wormhole (WH)
that may have a very distinctive GW signature. Specifically we focus
on  Lorentzian wormholes that are traversable  \cite{Morris:1988tu,Morris:1988cz,Visser:1995cc} and  constructed surgically from
Schwarzschild black hole (BH) spacetimes. We then consider a BH 
of mass $\Mbh $ falling into the wormhole of apparent mass
$\Mwh $, where we require $\Mbh < \Mwh $. For technical reasons we take the ratio
$\rho\equiv{M_{W\!H}}/{\Mbh} \gtrsim O(10)$, as will be discussed below. This range 
could be extended to larger $\rho$ at the expense of longer simulation run times, or
to smaller $\rho$ with further theoretical analysis and more careful approximations.  In Figure~\ref{fig:WormholeEmbed}, we show an embedding diagram for a wormhole, following \cite{Morris:1988cz} for how the embedding is done.

Let us begin with a brief review of the simplest surgical construction of a Schwarzschild 
wormhole \cite{Visser:1989kg,Visser:1989am}. (A variety of other types of wormholes that can be surgically constructed can be found in \cite{Visser:1989am}.) Starting with two Schwarzschild 
black holes of equivalent ADM mass $M=\Mwh$ \cite{Arnowitt:1962hi}, the line element is of the generic form:
\begin{equation}
ds^2 = -\left(1-\frac{2M}{r}\right) dt^2 + \frac{dr^2}{\left(1-\frac{2M}{r}\right)} +r^2(d{\theta}^2+{\sin}^2 \theta \, d{\phi}^2).
\end{equation}
We excise the interiors of both by removing everything
inside a sphere of radius of $a > 2M$, which creates a boundary for each BH and where $a$ will be determined below. Next
we identify the two boundaries, making sure the resulting manifold remains orientable. 
Finally we distribute exotic energy density at the boundary as required by the junction formalism \cite{Israel:1966rt}
to keep the Einstein equations satisfied everywhere. 
On converting to Gaussian normal coordinates for the static case, the metric for the complete space is: 
\begin{equation}
ds^2 = -\left(1-\frac{2\Mwh}{R(\eta)}\right) dt^2 + d{\eta}^2+{R(\eta)}^2(d{\theta}^2+{\sin}^2 \theta \, d{\phi}^2),
\end{equation}
where the Gaussian normal coordinate  $\eta$ is the proper radius and $R(\eta)$ satisfies
\begin{equation}
\frac{R}{\eta} =\pm \sqrt{1-\frac{2\Mwh}{R}}.
\end{equation}
The wormhole is composed of three regions: the exteriors of the Schwarzschild BHs where the stress--energy tensor is known (denoted here as {\it Universe~1} and {\it Universe~2}), and the {\it boundary}, where the surface stress--energy tensor is to be determined by assumption or experiment. $\eta$ is positive in Universe~1, negative in Universe~2, and zero on the boundary or throat.

The stress--energy tensor is
\begin{equation}
T^{\mu\nu} = S^{\mu\nu} \cdot \delta(\eta)
+T^{\mu\nu}_1\cdot\Theta(\eta)
+T^{\mu\nu}_{2}\cdot\Theta(-\eta),
\end{equation}
where $\Theta(\eta)$ is a step function. For Einstein gravity, the matter required on the wormhole throat invariably violates one or more energy conditions \cite{Hawking:1973uf}, depending on its equation of state. This can lead to instabilities in some cases~\cite{Visser:1989am,Visser:1989kg,Buniy:2005vh,Buniy:2006xf}.

As a simple example of a Schwarzschild wormhole, we envision material with equation of state $\sigma=\tau$ on the throat, where $\sigma$ is a negative surface energy density and $\tau$ is a negative pressure, i.e., surface tension. While this equation of state is similar to a negative energy density classical membrane, the matter need not be a membrane; it could consist of particles with  attractive interactions. Hence, the stress-energy tensor on the throat hypersurface where $\eta=0$  is
\begin{equation}
S^i{}_j = \left( \begin{matrix}\sigma&0&0\\
                            0&-\tau&0\\
                            0&0&-\tau\end{matrix} \right)_.
\end{equation}
The Einstein equations for $\sigma$ and $\tau$ on the throat lead to a throat radius $a=3 M = 3\Mwh$ \cite{Visser:1989kg}, identifying the mass $M$ with the effective wormhole mass $\Mwh$.

Now consider a black hole falling into a wormhole of the type just described. Since the WH looks just like another BH outside of a radius of $3\Mwh$,
the infall is indistinguishable from a BH-BH merger to an observer in Universe~1 until the BH reaches that radius. However, at that point the BH encounters the material on the throat. The interaction may be rather benign or not, depending on the properties of the material. Again we assume the simplest possibility: that there are only gravitational interactions between the throat matter and the infalling BH, and that the throat matter is made up of particles confined to the throat, but are otherwise interacting and free to move. Since the throat matter has negative energy density, it is repelled by the positive mass BH, causing it to simply move aside as the BH passes through. After the BH transit, the throat matter will redistribute itself in its original state with only minimal temporary disruption of the WH. The BH then passes through and spirals out the other end of the WH into Universe~2. Depending on the radial component of the BH's momentum, the BH can either move off to infinity in Universe~2 or be trapped in a series of transits through the WH.

What would we expect to observe from the above scenario? We assume the mouths of the wormhole in Universe~1 and Universe~2 are well separated so that an observed GW signal in Universe~1 will not overlap with a signal associated with the emergence of the BH in Universe~2, unless a shortcut is taken through the WH. The initial inspiral of the BH looks to an observer in Universe~1 like a typical BH-BH inspiral and merger with a normal GW chirp until the BH reaches the throat at $3\Mwh$. At that point the BH passes through the throat into Universe~2 and the gravitational wave signature in Universe~1 fades quickly as the BH is now radiating predominantly into Universe~2. An observer in Universe~2 sees a very strong signal that decreases in amplitude and frequency, an {\it anti-chirp}, until the BH reaches apogee and falls back into the WH.
The frequency reaches a maximum again at the throat as it passes from Universe~2 to 1, and the process repeats -- with chirps resulting from the inspiral BH trajectories and anti-chirps from outspiral BHs. With sufficient radial velocity, the can BH have such a highly eccentric orbit that GW emission near each apogee will be undetectable, reminiscent of {\it zoom-whirl} orbits seen in Extreme Mass Ratio Inspirals~\cite{Glampedakis2002}. If the BH is bound to the WH, it eventually settles near the throat of the WH with consequences depending on the properties of the unknown exotic matter on the throat.

Anti-chirps and gaps in the GW frequency spectrum are distinctive signatures that could be used to constrain the number of WHs in our Universe.  These GW signatures could be searched for in existing LIGO data by comparing with anti-chirp-like waveforms. While these constraints would only apply to a limited range of wormhole types -- namely traversable ones -- they could still have astrophysical and cosmological implications.

In the example that follows, let us assume the black hole mass is  $\Mbh= 5M_{\odot} $ and the effective mass of the wormhole is $\Mwh= 200M_{\odot}$. The throat is at radius $a=3\Mwh = 60 R_{BH}$, leading to a cross sectional area of the WH throat that is 3600 times as large as the area of the BH horizon. Thus it may not be unreasonable to consider a BH passing through the WH throat as a perturbation of the WH if the interaction with the exotic matter is not too strong; this would be the case if the exotic matter is composed of evenly distributed particles on the throat that only interact gravitationally with the BH ($\sigma < 0$). 
Under these constraints, the traversing BH is not expected to accrete exotic matter, as it is repelled by the positive mass BH\footnote{If exotic matter has different constraints, or react to additional forces, BH accretion must be reconsidered.}.


\section{Orbital Dynamics and Gravitational Waves  }

There is a rich suite of possible WH-BH orbits, even in the relatively simple regime in which $M_{\rm{WH}} \gg M_{\rm{BH}}$. For a Schwarzschild-like WH, the throat radius is at 1.5 times the Schwarzschild radius, $R_{s,WH}$ for a Schwarzschild BH of the same mass ---we stress, though, that there is no event horizon for the wormhole. For such unequal masses, the BH passing through the exotic matter at the throat is a minor perturbation to the matter and geometry of the system, which facilitates a post-Newtonian treatment to the dynamics.
We find that a bound BH will move through the WH many times in a damped oscillation, emitting energy into GWs on each side of the throat.  This results in the BH orbit decaying and settling on the throat of the WH. In this work we will explore non-eccentric orbits, which is the expected situation for binary systems due to the circularizing effects of the radiation reaction on the system  \cite{2014LRR....17....2B}. However, one could extend this approach to consider systems with eccentricity developed, for example, via the Kozai-Lidov mechanism \cite{Kozai:1962zz,1962P&SS....9..719L}.

Following the work of Diemer \& Smolarek \cite{DiemerSmolarek:2013mwv}, 
BH orbits are categorized as: \textit{bound orbits} (BO) - orbital motion is around the throat of the wormhole, never passing to the other side; \textit{two-world bound} (TWB) - orbits passing through the throat of the wormhole at least once; \textit{two-world escape} (TWE) -  orbits with enough energy that they go from radial infinity, pass through the wormhole, and outspiral to radial infinity on the other side~\footnote{It is an open question if these exist when GWs are considered.}; and \textit{escape orbits} (EO) which come in from radial infinity, swing around the wormhole and escape back out to radial infinity staying in a single universe.  In future work, we will consider a suite of trajectory topologies and their subsequent GW signatures. 

In this work, we consider what would be a circular, bound orbit (BO) in our Universe if the primary mass were a BH, but is a TWB orbit when the primary is a WH. Including gravitational radiation, the orbit loses energy and spirals into the wormhole throat, passes to the other Universe, and continues damped oscillations between spacetime regions, until it eventually settles to the throat.

As a first step to model the dynamics, we use the code ARCHAIN written by S.~Mikkola \citep{MikkolaMerritt2006, mikkola2008implementing} for integrating the two-body equations of motion up to 3.5~post-Newtonian (PN) order, including gravitational radiation at 2.5~PN.  We treat the wormhole as effectively a large mass with a smaller black hole orbiting it. Using the post-Newtonian approximation is not only a convenient choice for explicitly solving for the orbit, but our mass ratio of 40:1 makes the problem computationally intractable for current numerical relativity simulations which have calculated ratios only as large as 18:1 with difficulty~\cite{PhysRevD.93.044006,2016CQGra..33t4001J}. One might be concerned with following the trajectory up to and through the wormhole, as PN approximations are thought to breakdown close to BH-BH mergers~\cite{PhysRevD.81.064004}; here, though, the orbital inspiral stops effectively at $ 1.5 R_{s,WH}$, as it traverses to a new Universe, which mitigates some of the mismatch between NR and PN-calculated dynamics.

Table~\ref{tab:WHBHICs} presents initial conditions for our near circular orbit. Included are the WH and BH center-of-mass positions and velocities in the $xy$-plane for the starting position near $x_{BH}\sim 20$ (in units of $R_{s,WH}$) which is integrated with a coarse step-size.  We take the positions and velocities from that trajectory near $x_{BH}\sim 7$ as the initial conditions to calculate a more finely-resolved trajectory.  We work in the center-of-mass coordinates in the figures that follow, with a mass ratio of 40:1 for the $\Mwh:\Mbh$. We use units such that $2\, \Mwh = R_{s,WH} = 1$, the gravitational constant $G=1$, and the speed-of-light $c=1$.  In SI units, one finds the unit of length is $U_l=5.9065\times 10^{5}\, \textrm{m}$, and of mass $U_m=400\, M_\odot = 7.9536\times 10^{32}\, \textrm{kg}$, and finally the unit of time is $U_t=1.9702\times 10^{-3}\, \textrm{s}$.  The wormhole has an effective mass of $200\, M_\odot$ while the black hole has mass $5\, M_\odot$, or in these units, $0.5$ and $0.0125$, respectively.  All motion is in the $xy$-plane, and neither the wormhole nor the black hole has spin.

\begin{table*}[ht]
    \centering
    \caption{\footnotesize{Initial conditions for the Wormhole and Black Hole, showing the center-of-mass positions and velocities in the $xy$-plane.  
    The first row denotes the initial conditions for a coarsely integrated trajectory used to find the position and velocities in the second row, which are used to start the finely integrated trajectory used for the GW calculation.}}
    \label{tab:WHBHICs}

    \begin{tabular}{|c|c|c|c||c|c|c|c||c|}
    \hline
    $x_{WH}$ & $y_{WH}$ & $v_{x,WH}$ & $v_{y,WH}$ & $x_{BH}$ & $y_{BH}$ & $v_{x,BH}$ & $v_{y,BH}$ & step size \\ \hline
    -0.4878 & 0 & 0 & -3.7611e-3 & 19.512 & 0 & 0 & 0.15045 & 4.0 \\
    -0.17973 & 3.4144e-4 & -7.3384e-06 & -5.8446e-3 & 7.1894 & -0.013658 & 2.9353e-4 & 0.23379 & 0.0001 \\ \hline
    \end{tabular}
\end{table*}

\begin{figure}[ht]
	\includegraphics[width=0.5\textwidth]{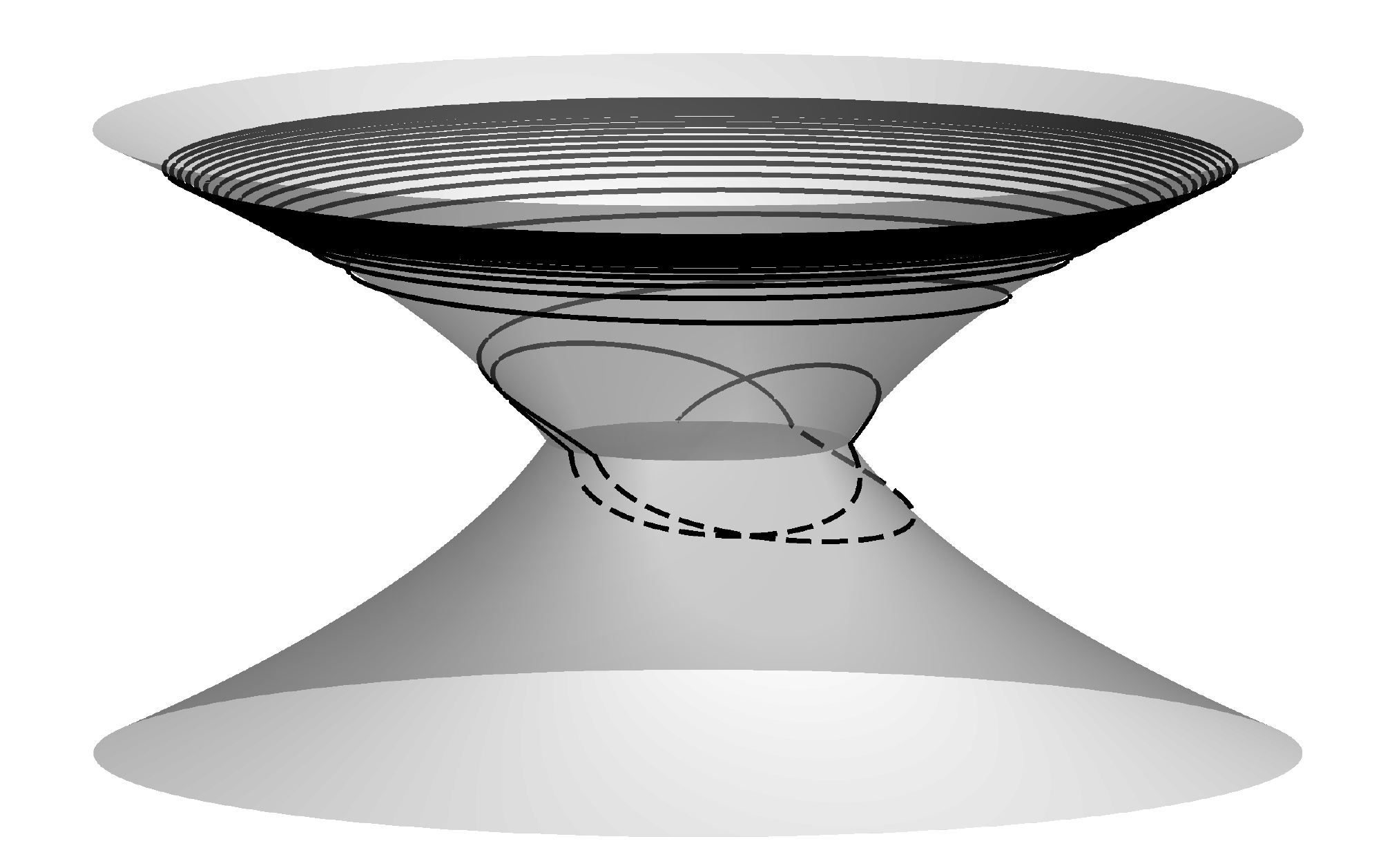}
	\caption{\footnotesize{Embedding diagram for an orbit following Diemer and Smolarek\cite{DiemerSmolarek:2013mwv}. The orbit of a black hole with mass $5~M_\odot$ around and through a wormhole with an effective mass $200~M_\odot$.  Without GWs this would be a bound, circular orbit in Universe~1 (top half).  Initial position is at the top of the hyperboloid; the inspiral of the BH can be seen.  As it passes the throat into Universe~2 (bottom half), we indicate orbits in Universe~2 with dashed lines. Five ``passes" or ``petals" are shown, three in Universe~1 and two (dashed) in Universe~2.  
	} 	}
	\label{fig:BHorbitsEmbed}
	\vspace{-3mm}
\end{figure}

In Figure~\ref{fig:BHorbitsEmbed}, we show the initial inspiral and four passes through the throat of the wormhole in an embedding diagram, following that of \cite{DiemerSmolarek:2013mwv}.  The upper half of the hyperboloid is Universe~1, it contains the initial position, and the lower half is Universe~2.  We show trajectories in Universe~2 as dashed, those in 1 as solid lines.

\begin{figure}[ht]
	\includegraphics[width=0.5\textwidth]{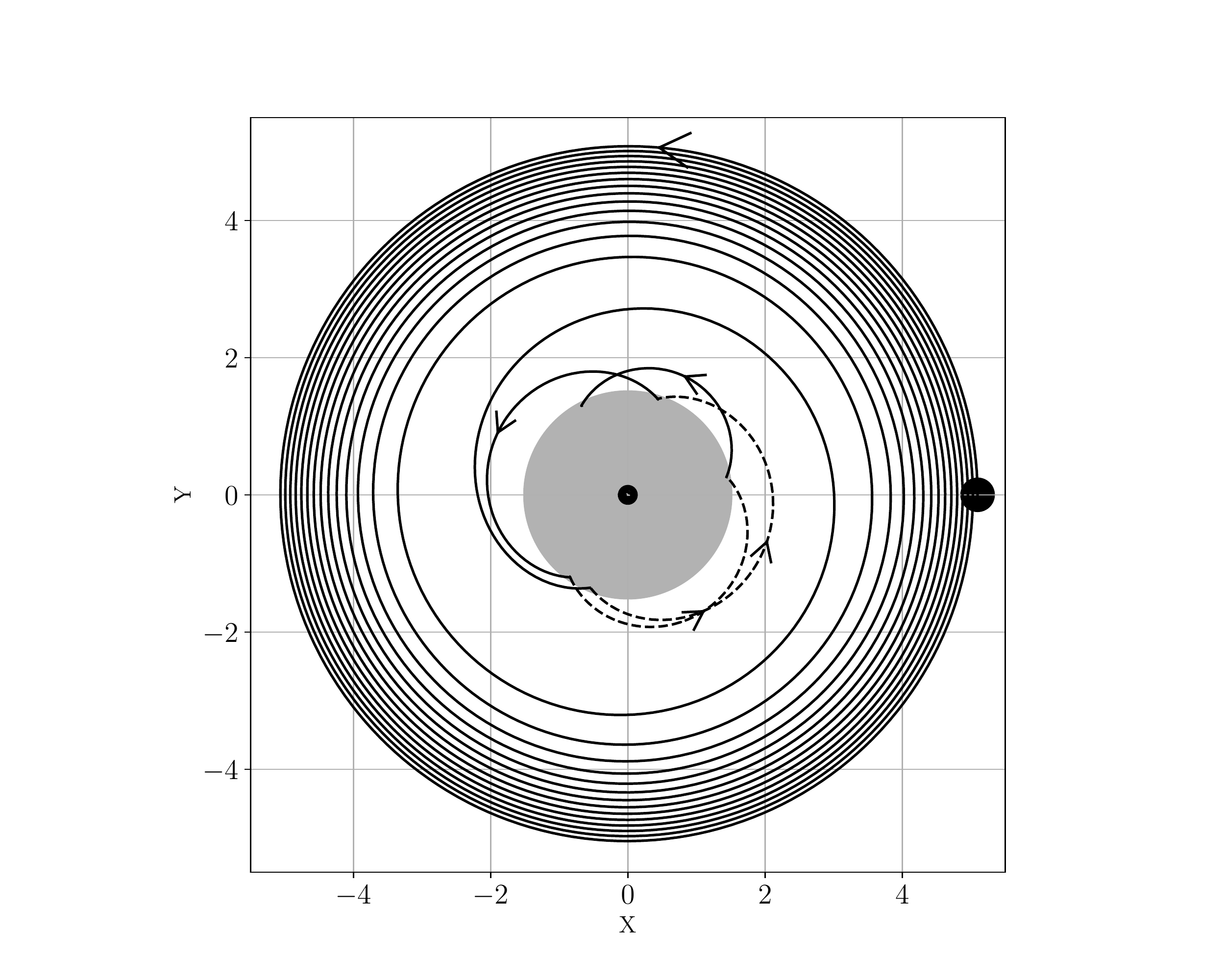}
	\caption{\footnotesize{The orbits of a black hole with mass $5~M_\odot$ around and through a wormhole with an effective mass $200~M_\odot$, same as in Figure~\ref{fig:BHorbitsEmbed} but in $xy$-(physical) space.  The orbit starts in Universe~1 and traverses to 2, then back in 1, etc. The black circle marks the initial position and the circulation of the orbits are counter-clockwise. The solid gray circle is approximately the wormhole throat--the center of the wormhole executes the small orbit indicated by the black dot around the origin. Orbits in Universe~2 are shown dashed.  
	} 	}
	\label{fig:BHorbits}
	\vspace{-3mm}
\end{figure}

\begin{figure}[ht]
	\includegraphics[width=0.5\textwidth]{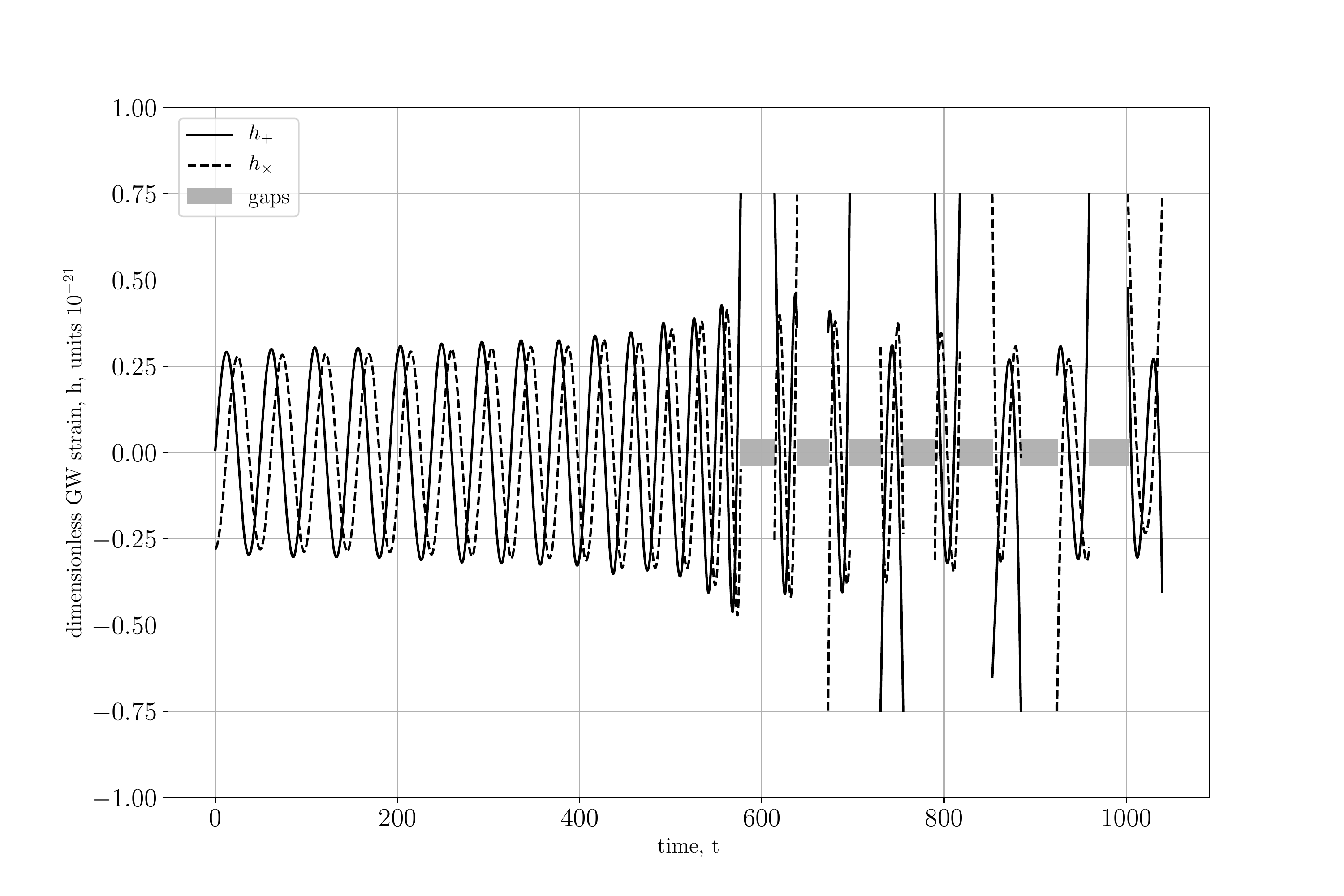}
	\caption{\footnotesize{The dimensionless gravitational strain measured for the wormhole-black hole binary shown in Figure~\ref{fig:BHorbits} at a distance of $500~\textrm{Mpc}$ and at an angle of $\phi=\pi/4$ and $\theta=\pi/4$ relative to the binary orbital plane.  There is some difficulty working with these discontinuous petals, so the endpoints are not considered accurate.  The wormhole mass is $200~M_\odot$ and the black hole is $5~M_\odot$.  The estimated gravitational wave frequency of this example is approximately  $14~\textrm{Hz}$ near $t=0$.  Smaller total mass would raise the frequency.  It is likely the gaps between bursts would fill with GW leakage from the other Universe, but this involves the exotic matter interacting with the GWs from Universe~2---here we tacitly assume those GWs are blocked completely.
	}}
	\label{fig:GWcalc}
	\vspace{-3mm}
\end{figure}

The part of the orbit shown in Figure~\ref{fig:BHorbits} corresponding to Universe~1 is used to calculate the gravitational strain shown in Figure~\ref{fig:GWcalc}.  The orbits show an interesting property, that the black holes will settle down into a position on the throat of the wormhole; presumably this would also hold for any matter with positive mass.  Asymptotically the angular momentum will radiate away, so the object stops orbiting and is at a fixed position on the throat sphere.  The wormhole could potentially collect a set of compact objects on its throat, the observational and theoretical consequences of such has yet to be fully explored.

We calculate the gravitational waves using the mass quadrupole formula given in \cite{maggiore2008gravitational} (see Equation~(3.72) there).  Since the strain $h\sim d^2 Q_{ij}/dt^2$, where $Q_{ij}$ is the mass quadrupole moment of the binary, we smooth the orbits using a 4th order spline to have a smooth second derivative.  This still gives rather poor behavior at the discontinuous endpoints of the orbit as it passes through the throat, see Figure~\ref{fig:GWcalc}, but acceptable behavior elsewhere.

We assumed a distance of 500~Mpc and an inclination of the binary orbital plane of $\phi=\theta=\pi/4$.  The two polarizations are shown in Figure~\ref{fig:GWcalc}, with plus-polarization $h_+$ shown as solid, and the cross-polarization $h_\times$ as dashed.  Using the above units one finds the inspiral to have a frequency of approximately $14~\textrm{Hz}$ near $t=0$. This is in the frequency band anticipated for future 3-G detectors~\cite{2019:3GWhitePaper}. 
We estimate the signal-to-noise for the waveform in Figure~\ref{fig:GWcalc}, with sky and polarization averaging~\citep{maggiore2008gravitational} to be 6 for Advanced-LIGO and 460 for the third generation gravitational wave detector Cosmic Explorer\citep{2019BAAS...51g..35R,CosmicExploreWebsite}. This scales inversely with the total mass, so a less massive system could be high enough frequency to be observed by LIGO.  In the figure, we use gray bars to emphasize the gaps in the GW waveform.  Generically, one can see that for a wormhole, a somewhat normal inspiral occurs and is reflected in the GW signature.  However, the BH does not merge but passes through the throat, creating a gap in the waveform earlier than it would had the WH been a BH.  In the case shown, the inspiral is followed by about one cycle bursts of GWs of slightly decreasing amplitude.

\section{Summary and Discussion}

Gravitational wave astronomy will continue to expand its sensitivity, eventually producing a census of massive binary inspirals extending to high redshifts that could encompass even the first generation of seed BH-BH mergers \cite[e.g.][]{Yu:2017zgi, ColpiWP}.  With this increased sensitivity, the possibility of constraining (or detecting!) exotic phenomena via GW signals has arrived. In the present work we have examined a novel anti-chirp signature arising from a bound binary traversable wormhole system where the bound object emerges through the wormhole throat into our region of spacetime. Note that that since the WH throat is at a larger radius than the event horizon of a BH of the same mass, a WH-BH interaction would be {\em distinguishable} even during an inspiral, because of this the signal would drop off at a lower frequency than is expected for the same mass BH-BH merger.  Also if there is obviously no ringdown GW signal from the final, merged BH, and one would expect it, this would also be evidence for something other than a BH-BH merger.

We have been assuming near circular orbits for BHs near WHs, but this does not always have to be the case. For example, Kozai \cite{Kozai:1962zz} and Lidov \cite{1962P&SS....9..719L} have shown that three body effects can give non-zero eccentricities to orbits that are initially circular. So, even though merger orbits tend to circularize due to the effects of GW radiation, if there is another body in the vicinity, the Kozai-Lidov mechanism could lead to a merger that is eccentric as a BH crosses a wormhole throat. Overall this would imply a higher chance of mergers  where the BH starting in Universe~1 either escapes into Universe~2 or starts off with an eccentric bound orbit in Universe~2. This is just one example of beyond circular orbit signal possibilities one could expect for BH-WH mergers.

Wormholes offer numerous intriguing scenarios for astronomy and cosmology.
They have been suggested as possible sources of fast radio burst \cite{Torres:1998xd} and other phenomena. They have the potential to  explain observations that have not yet been fully understood, e.g., the recent LIGO-VIRGO event S190521g observations  \cite{Graham:2020gwr}.
They also provide another form of compact object as an alternative to a supermassive black hole, e.g., as the only possibility for Sgr $A^*$ leading to a consistent explanation for the orbits of stars in its neighborhood \cite{Dai:2019mse}. Their discovery could imply the possible existence of exotic matter \cite{Morris:1988tu,Morris:1988cz,Hochberg:1991tz}---though wormholes in a modified gravity scenario are possible without exotic matter \cite{Duplessis:2015xva,Dent:2016efw}. Wormholes would also allow shortcuts through our Universe which has implications for cosmology \cite{Hochberg:1992du}.

This initial work posits the unique gravitational signature of a potential WH-BH interaction, but it merely scratches the surface of this domain. For example, we considered a Schwarzschild-like traversable wormhole with a rather simplified choice of exotic matter, but a spinning WH solution is more astrophysically relevant. In addition to mapping more eccentric orbits, future avenues include the effect of higher-order modes on the outspiral waveforms; astrophysical implications of such WH binaries, which can place constraints on WH demographics; and an exploration of the potential multi-messenger signals of WH-BH systems.

\begin{acknowledgments}

{\bf\emph{Acknowledgments-}} The work of TWK is supported by U.S. DoE grant number DE-SC-0019235. JBD acknowledges support from the National Science Foundation under Grant No. NSF PHY-1820801. JBD would also like to thank the Mitchell Institute for Fundamental Physics and Astronomy at Texas A\&M University for their generous hospitality during various stages of this work.
The authors would like to thank Dr.~Karan Jani for valuable discussions.
\end{acknowledgments}

\bibliography{whgw}

\begin{thebibliography}{48}%
\makeatletter
\providecommand \@ifxundefined [1]{%
 \@ifx{#1\undefined}
}%
\providecommand \@ifnum [1]{%
 \ifnum #1\expandafter \@firstoftwo
 \else \expandafter \@secondoftwo
 \fi
}%
\providecommand \@ifx [1]{%
 \ifx #1\expandafter \@firstoftwo
 \else \expandafter \@secondoftwo
 \fi
}%
\providecommand \natexlab [1]{#1}%
\providecommand \enquote  [1]{``#1''}%
\providecommand \bibnamefont  [1]{#1}%
\providecommand \bibfnamefont [1]{#1}%
\providecommand \citenamefont [1]{#1}%
\providecommand \href@noop [0]{\@secondoftwo}%
\providecommand \href [0]{\begingroup \@sanitize@url \@href}%
\providecommand \@href[1]{\@@startlink{#1}\@@href}%
\providecommand \@@href[1]{\endgroup#1\@@endlink}%
\providecommand \@sanitize@url [0]{\catcode `\\12\catcode `\$12\catcode
  `\&12\catcode `\#12\catcode `\^12\catcode `\_12\catcode `\%12\relax}%
\providecommand \@@startlink[1]{}%
\providecommand \@@endlink[0]{}%
\providecommand \url  [0]{\begingroup\@sanitize@url \@url }%
\providecommand \@url [1]{\endgroup\@href {#1}{\urlprefix }}%
\providecommand \urlprefix  [0]{URL }%
\providecommand \Eprint [0]{\href }%
\providecommand \doibase [0]{http://dx.doi.org/}%
\providecommand \selectlanguage [0]{\@gobble}%
\providecommand \bibinfo  [0]{\@secondoftwo}%
\providecommand \bibfield  [0]{\@secondoftwo}%
\providecommand \translation [1]{[#1]}%
\providecommand \BibitemOpen [0]{}%
\providecommand \bibitemStop [0]{}%
\providecommand \bibitemNoStop [0]{.\EOS\space}%
\providecommand \EOS [0]{\spacefactor3000\relax}%
\providecommand \BibitemShut  [1]{\csname bibitem#1\endcsname}%
\let\auto@bib@innerbib\@empty
\bibitem [{\citenamefont {Abbott}\ \emph {et~al.}(2016)\citenamefont {Abbott}
  \emph {et~al.}}]{Abbott:2016blz}%
  \BibitemOpen
  \bibfield  {author} {\bibinfo {author} {\bibfnamefont {B.~P.}\ \bibnamefont
  {Abbott}} \emph {et~al.} (\bibinfo {collaboration} {LIGO Scientific,
  Virgo}),\ }\href {\doibase 10.1103/PhysRevLett.116.061102} {\bibfield
  {journal} {\bibinfo  {journal} {Phys. Rev. Lett.}\ }\textbf {\bibinfo
  {volume} {116}},\ \bibinfo {pages} {061102} (\bibinfo {year} {2016})},\
  \Eprint {http://arxiv.org/abs/1602.03837} {arXiv:1602.03837 [gr-qc]}
  \BibitemShut {NoStop}%
\bibitem [{\citenamefont {Abbott}\ \emph {et~al.}(2017)\citenamefont {Abbott}
  \emph {et~al.}}]{TheLIGOScientific:2017qsa}%
  \BibitemOpen
  \bibfield  {author} {\bibinfo {author} {\bibfnamefont {B.~P.}\ \bibnamefont
  {Abbott}} \emph {et~al.} (\bibinfo {collaboration} {LIGO Scientific,
  Virgo}),\ }\href {\doibase 10.1103/PhysRevLett.119.161101} {\bibfield
  {journal} {\bibinfo  {journal} {Phys. Rev. Lett.}\ }\textbf {\bibinfo
  {volume} {119}},\ \bibinfo {pages} {161101} (\bibinfo {year} {2017})},\
  \Eprint {http://arxiv.org/abs/1710.05832} {arXiv:1710.05832 [gr-qc]}
  \BibitemShut {NoStop}%
\bibitem [{\citenamefont {Abbott}\ \emph {et~al.}(2020)\citenamefont {Abbott}
  \emph {et~al.}}]{Abbott:2020uma}%
  \BibitemOpen
  \bibfield  {author} {\bibinfo {author} {\bibfnamefont {B.~P.}\ \bibnamefont
  {Abbott}} \emph {et~al.} (\bibinfo {collaboration} {LIGO Scientific,
  Virgo}),\ }\href {\doibase 10.3847/2041-8213/ab75f5} {\bibfield  {journal}
  {\bibinfo  {journal} {Astrophys. J. Lett.}\ }\textbf {\bibinfo {volume}
  {892}},\ \bibinfo {pages} {L3} (\bibinfo {year} {2020})},\ \Eprint
  {http://arxiv.org/abs/2001.01761} {arXiv:2001.01761 [astro-ph.HE]}
  \BibitemShut {NoStop}%
\bibitem [{\citenamefont {Abbott}\ \emph {et~al.}(2019)\citenamefont {Abbott}
  \emph {et~al.}}]{PhysRevX.9.031040}%
  \BibitemOpen
  \bibfield  {author} {\bibinfo {author} {\bibfnamefont {B.~P.}\ \bibnamefont
  {Abbott}} \emph {et~al.} (\bibinfo {collaboration} {LIGO Scientific,
  Virgo}),\ }\href {\doibase 10.1103/PhysRevX.9.031040} {\bibfield  {journal}
  {\bibinfo  {journal} {Phys. Rev. X}\ }\textbf {\bibinfo {volume} {9}},\
  \bibinfo {pages} {031040} (\bibinfo {year} {2019})}\BibitemShut {NoStop}%
\bibitem [{\citenamefont {Vilenkin}\ and\ \citenamefont
  {Shellard}(2000)}]{Vilenkin:2000jqa}%
  \BibitemOpen
  \bibfield  {author} {\bibinfo {author} {\bibfnamefont {A.}~\bibnamefont
  {Vilenkin}}\ and\ \bibinfo {author} {\bibfnamefont {E.~P.~S.}\ \bibnamefont
  {Shellard}},\ }\href
  {http://www.cambridge.org/mw/academic/subjects/physics/theoretical-physics-and-mathematical-physics/cosmic-strings-and-other-topological-defects?format=PB}
  {\emph {\bibinfo {title} {{Cosmic Strings and Other Topological Defects}}}}\
  (\bibinfo  {publisher} {Cambridge University Press},\ \bibinfo {year}
  {2000})\BibitemShut {NoStop}%
\bibitem [{\citenamefont {Wong}\ \emph {et~al.}(2019)\citenamefont {Wong},
  \citenamefont {Berti}, \citenamefont {Gabella},\ and\ \citenamefont
  {Holley-Bockelmann}}]{Wong:2018amf}%
  \BibitemOpen
  \bibfield  {author} {\bibinfo {author} {\bibfnamefont {K.~W.~K.}\
  \bibnamefont {Wong}}, \bibinfo {author} {\bibfnamefont {E.}~\bibnamefont
  {Berti}}, \bibinfo {author} {\bibfnamefont {W.~E.}\ \bibnamefont {Gabella}},
  \ and\ \bibinfo {author} {\bibfnamefont {K.}~\bibnamefont
  {Holley-Bockelmann}},\ }\href {\doibase 10.1093/mnrasl/sly208} {\bibfield
  {journal} {\bibinfo  {journal} {Mon. Not. Roy. Astron. Soc.}\ }\textbf
  {\bibinfo {volume} {483}},\ \bibinfo {pages} {L33} (\bibinfo {year}
  {2019})},\ \Eprint {http://arxiv.org/abs/1808.07055} {arXiv:1808.07055
  [astro-ph.EP]} \BibitemShut {NoStop}%
\bibitem [{\citenamefont {Wheeler}(1955)}]{Wheeler:1955zz}%
  \BibitemOpen
  \bibfield  {author} {\bibinfo {author} {\bibfnamefont {J.~A.}\ \bibnamefont
  {Wheeler}},\ }\href {\doibase 10.1103/PhysRev.97.511} {\bibfield  {journal}
  {\bibinfo  {journal} {Phys. Rev.}\ }\textbf {\bibinfo {volume} {97}},\
  \bibinfo {pages} {511} (\bibinfo {year} {1955})}\BibitemShut {NoStop}%
\bibitem [{\citenamefont {Wheeler}(1962)}]{Wheeler:1962}%
  \BibitemOpen
  \bibfield  {author} {\bibinfo {author} {\bibfnamefont {J.}~\bibnamefont
  {Wheeler}},\ }\href@noop {} {\emph {\bibinfo {title} {{Geometrodynamics}}}}\
  (\bibinfo  {publisher} {Academic Press},\ \bibinfo {year} {1962})\BibitemShut
  {NoStop}%
\bibitem [{\citenamefont {Cardoso}\ \emph
  {et~al.}(2016{\natexlab{a}})\citenamefont {Cardoso}, \citenamefont
  {Franzin},\ and\ \citenamefont {Pani}}]{Cardoso:2016rao}%
  \BibitemOpen
  \bibfield  {author} {\bibinfo {author} {\bibfnamefont {V.}~\bibnamefont
  {Cardoso}}, \bibinfo {author} {\bibfnamefont {E.}~\bibnamefont {Franzin}}, \
  and\ \bibinfo {author} {\bibfnamefont {P.}~\bibnamefont {Pani}},\ }\href
  {\doibase 10.1103/PhysRevLett.117.089902, 10.1103/PhysRevLett.116.171101}
  {\bibfield  {journal} {\bibinfo  {journal} {Phys. Rev. Lett.}\ }\textbf
  {\bibinfo {volume} {116}},\ \bibinfo {pages} {171101} (\bibinfo {year}
  {2016}{\natexlab{a}})},\ \bibinfo {note} {[Erratum: Phys. Rev.
  Lett.117,no.8,089902(2016)]},\ \Eprint {http://arxiv.org/abs/1602.07309}
  {arXiv:1602.07309 [gr-qc]} \BibitemShut {NoStop}%
\bibitem [{\citenamefont {Cardoso}\ \emph
  {et~al.}(2016{\natexlab{b}})\citenamefont {Cardoso}, \citenamefont {Hopper},
  \citenamefont {Macedo}, \citenamefont {Palenzuela},\ and\ \citenamefont
  {Pani}}]{Cardoso:2016oxy}%
  \BibitemOpen
  \bibfield  {author} {\bibinfo {author} {\bibfnamefont {V.}~\bibnamefont
  {Cardoso}}, \bibinfo {author} {\bibfnamefont {S.}~\bibnamefont {Hopper}},
  \bibinfo {author} {\bibfnamefont {C.~F.~B.}\ \bibnamefont {Macedo}}, \bibinfo
  {author} {\bibfnamefont {C.}~\bibnamefont {Palenzuela}}, \ and\ \bibinfo
  {author} {\bibfnamefont {P.}~\bibnamefont {Pani}},\ }\href {\doibase
  10.1103/PhysRevD.94.084031} {\bibfield  {journal} {\bibinfo  {journal} {Phys.
  Rev.}\ }\textbf {\bibinfo {volume} {D94}},\ \bibinfo {pages} {084031}
  (\bibinfo {year} {2016}{\natexlab{b}})},\ \Eprint
  {http://arxiv.org/abs/1608.08637} {arXiv:1608.08637 [gr-qc]} \BibitemShut
  {NoStop}%
\bibitem [{\citenamefont {Bueno}\ \emph {et~al.}(2018)\citenamefont {Bueno},
  \citenamefont {Cano}, \citenamefont {Goelen}, \citenamefont {Hertog},\ and\
  \citenamefont {Vercnocke}}]{Bueno:2017hyj}%
  \BibitemOpen
  \bibfield  {author} {\bibinfo {author} {\bibfnamefont {P.}~\bibnamefont
  {Bueno}}, \bibinfo {author} {\bibfnamefont {P.~A.}\ \bibnamefont {Cano}},
  \bibinfo {author} {\bibfnamefont {F.}~\bibnamefont {Goelen}}, \bibinfo
  {author} {\bibfnamefont {T.}~\bibnamefont {Hertog}}, \ and\ \bibinfo {author}
  {\bibfnamefont {B.}~\bibnamefont {Vercnocke}},\ }\href {\doibase
  10.1103/PhysRevD.97.024040} {\bibfield  {journal} {\bibinfo  {journal} {Phys.
  Rev.}\ }\textbf {\bibinfo {volume} {D97}},\ \bibinfo {pages} {024040}
  (\bibinfo {year} {2018})},\ \Eprint {http://arxiv.org/abs/1711.00391}
  {arXiv:1711.00391 [gr-qc]} \BibitemShut {NoStop}%
\bibitem [{\citenamefont {Aneesh}\ \emph {et~al.}(2018)\citenamefont {Aneesh},
  \citenamefont {Bose},\ and\ \citenamefont {Kar}}]{Aneesh:2018hlp}%
  \BibitemOpen
  \bibfield  {author} {\bibinfo {author} {\bibfnamefont {S.}~\bibnamefont
  {Aneesh}}, \bibinfo {author} {\bibfnamefont {S.}~\bibnamefont {Bose}}, \ and\
  \bibinfo {author} {\bibfnamefont {S.}~\bibnamefont {Kar}},\ }\href {\doibase
  10.1103/PhysRevD.97.124004} {\bibfield  {journal} {\bibinfo  {journal} {Phys.
  Rev.}\ }\textbf {\bibinfo {volume} {D97}},\ \bibinfo {pages} {124004}
  (\bibinfo {year} {2018})},\ \Eprint {http://arxiv.org/abs/1803.10204}
  {arXiv:1803.10204 [gr-qc]} \BibitemShut {NoStop}%
\bibitem [{\citenamefont {Abedi}\ and\ \citenamefont
  {Afshordi}(2019)}]{Abedi:2018npz}%
  \BibitemOpen
  \bibfield  {author} {\bibinfo {author} {\bibfnamefont {J.}~\bibnamefont
  {Abedi}}\ and\ \bibinfo {author} {\bibfnamefont {N.}~\bibnamefont
  {Afshordi}},\ }\href {\doibase 10.1088/1475-7516/2019/11/010} {\bibfield
  {journal} {\bibinfo  {journal} {JCAP}\ }\textbf {\bibinfo {volume} {1911}},\
  \bibinfo {pages} {010} (\bibinfo {year} {2019})},\ \Eprint
  {http://arxiv.org/abs/1803.10454} {arXiv:1803.10454 [gr-qc]} \BibitemShut
  {NoStop}%
\bibitem [{\citenamefont {Abedi}\ and\ \citenamefont
  {Afshordi}(2020)}]{Abedi:2020sgg}%
  \BibitemOpen
  \bibfield  {author} {\bibinfo {author} {\bibfnamefont {J.}~\bibnamefont
  {Abedi}}\ and\ \bibinfo {author} {\bibfnamefont {N.}~\bibnamefont
  {Afshordi}},\ }\href@noop {} {\bibfield  {journal} {\bibinfo  {journal}
  {{arXiv e-prints}}\ ,\ \bibinfo {eid} {{arXiv:2001.00821}}} (\bibinfo {year}
  {2020})},\ \Eprint {http://arxiv.org/abs/2001.00821} {arXiv:2001.00821
  [gr-qc]} \BibitemShut {NoStop}%
\bibitem [{\citenamefont {Kirillov}\ \emph {et~al.}(2020)\citenamefont
  {Kirillov}, \citenamefont {Savelova},\ and\ \citenamefont
  {Lecian}}]{Kirillov:2020cre}%
  \BibitemOpen
  \bibfield  {author} {\bibinfo {author} {\bibfnamefont {A.~A.}\ \bibnamefont
  {Kirillov}}, \bibinfo {author} {\bibfnamefont {E.~P.}\ \bibnamefont
  {Savelova}}, \ and\ \bibinfo {author} {\bibfnamefont {O.~M.}\ \bibnamefont
  {Lecian}},\ }\href@noop {} {\  (\bibinfo {year} {2020})},\ \Eprint
  {http://arxiv.org/abs/2003.13127} {arXiv:2003.13127 [gr-qc]} \BibitemShut
  {NoStop}%
\bibitem [{\citenamefont {Morris}\ \emph {et~al.}(1988)\citenamefont {Morris},
  \citenamefont {Thorne},\ and\ \citenamefont {Yurtsever}}]{Morris:1988tu}%
  \BibitemOpen
  \bibfield  {author} {\bibinfo {author} {\bibfnamefont {M.~S.}\ \bibnamefont
  {Morris}}, \bibinfo {author} {\bibfnamefont {K.~S.}\ \bibnamefont {Thorne}},
  \ and\ \bibinfo {author} {\bibfnamefont {U.}~\bibnamefont {Yurtsever}},\
  }\href {\doibase 10.1103/PhysRevLett.61.1446} {\bibfield  {journal} {\bibinfo
   {journal} {Phys. Rev. Lett.}\ }\textbf {\bibinfo {volume} {61}},\ \bibinfo
  {pages} {1446} (\bibinfo {year} {1988})}\BibitemShut {NoStop}%
\bibitem [{\citenamefont {Morris}\ and\ \citenamefont
  {Thorne}(1988)}]{Morris:1988cz}%
  \BibitemOpen
  \bibfield  {author} {\bibinfo {author} {\bibfnamefont {M.~S.}\ \bibnamefont
  {Morris}}\ and\ \bibinfo {author} {\bibfnamefont {K.~S.}\ \bibnamefont
  {Thorne}},\ }\href {\doibase 10.1119/1.15620} {\bibfield  {journal} {\bibinfo
   {journal} {Am. J. Phys.}\ }\textbf {\bibinfo {volume} {56}},\ \bibinfo
  {pages} {395} (\bibinfo {year} {1988})}\BibitemShut {NoStop}%
\bibitem [{\citenamefont {Visser}(1995)}]{Visser:1995cc}%
  \BibitemOpen
  \bibfield  {author} {\bibinfo {author} {\bibfnamefont {M.}~\bibnamefont
  {Visser}},\ }\href@noop {} {\emph {\bibinfo {title} {{Lorentzian wormholes:
  From Einstein to Hawking}}}}\ (\bibinfo  {publisher} {AIP Press, Woodbury,
  New York},\ \bibinfo {year} {1995})\BibitemShut {NoStop}%
\bibitem [{\citenamefont {Visser}(1989)}]{Visser:1989kg}%
  \BibitemOpen
  \bibfield  {author} {\bibinfo {author} {\bibfnamefont {M.}~\bibnamefont
  {Visser}},\ }\href {\doibase 10.1016/0550-3213(89)90100-4} {\bibfield
  {journal} {\bibinfo  {journal} {Nucl. Phys.}\ }\textbf {\bibinfo {volume}
  {B328}},\ \bibinfo {pages} {203} (\bibinfo {year} {1989})},\ \Eprint
  {http://arxiv.org/abs/0809.0927} {arXiv:0809.0927 [gr-qc]} \BibitemShut
  {NoStop}%
\bibitem [{\citenamefont {Visser}(1990)}]{Visser:1989am}%
  \BibitemOpen
  \bibfield  {author} {\bibinfo {author} {\bibfnamefont {M.}~\bibnamefont
  {Visser}},\ }\href {\doibase 10.1016/0370-2693(90)91588-3} {\bibfield
  {journal} {\bibinfo  {journal} {Phys. Lett.}\ }\textbf {\bibinfo {volume}
  {B242}},\ \bibinfo {pages} {24} (\bibinfo {year} {1990})}\BibitemShut
  {NoStop}%
\bibitem [{\citenamefont {Arnowitt}\ \emph {et~al.}(2008)\citenamefont
  {Arnowitt}, \citenamefont {Deser},\ and\ \citenamefont
  {Misner}}]{Arnowitt:1962hi}%
  \BibitemOpen
  \bibfield  {author} {\bibinfo {author} {\bibfnamefont {R.~L.}\ \bibnamefont
  {Arnowitt}}, \bibinfo {author} {\bibfnamefont {S.}~\bibnamefont {Deser}}, \
  and\ \bibinfo {author} {\bibfnamefont {C.~W.}\ \bibnamefont {Misner}},\
  }\href {\doibase 10.1007/s10714-008-0661-1} {\bibfield  {journal} {\bibinfo
  {journal} {Gen. Rel. Grav.}\ }\textbf {\bibinfo {volume} {40}},\ \bibinfo
  {pages} {1997} (\bibinfo {year} {2008})},\ \Eprint
  {http://arxiv.org/abs/gr-qc/0405109} {arXiv:gr-qc/0405109 [gr-qc]}
  \BibitemShut {NoStop}%
\bibitem [{\citenamefont {Israel}(1966)}]{Israel:1966rt}%
  \BibitemOpen
  \bibfield  {author} {\bibinfo {author} {\bibfnamefont {W.}~\bibnamefont
  {Israel}},\ }\href {\doibase 10.1007/BF02710419, 10.1007/BF02712210}
  {\bibfield  {journal} {\bibinfo  {journal} {Nuovo Cim.}\ }\textbf {\bibinfo
  {volume} {B44S10}},\ \bibinfo {pages} {1} (\bibinfo {year} {1966})},\
  \bibinfo {note} {[Erratum: Nuovo Cim.B48,463(1967); Nuovo
  Cim.B44,1(1966)]}\BibitemShut {NoStop}%
\bibitem [{\citenamefont {Hawking}\ and\ \citenamefont
  {Ellis}(2011)}]{Hawking:1973uf}%
  \BibitemOpen
  \bibfield  {author} {\bibinfo {author} {\bibfnamefont {S.~W.}\ \bibnamefont
  {Hawking}}\ and\ \bibinfo {author} {\bibfnamefont {G.~F.~R.}\ \bibnamefont
  {Ellis}},\ }\href {\doibase 10.1017/CBO9780511524646} {\emph {\bibinfo
  {title} {{The Large Scale Structure of Space-Time}}}},\ Cambridge Monographs
  on Mathematical Physics\ (\bibinfo  {publisher} {Cambridge University
  Press},\ \bibinfo {year} {2011})\BibitemShut {NoStop}%
\bibitem [{\citenamefont {Buniy}\ and\ \citenamefont
  {Hsu}(2006)}]{Buniy:2005vh}%
  \BibitemOpen
  \bibfield  {author} {\bibinfo {author} {\bibfnamefont {R.~V.}\ \bibnamefont
  {Buniy}}\ and\ \bibinfo {author} {\bibfnamefont {S.~D.~H.}\ \bibnamefont
  {Hsu}},\ }\href {\doibase 10.1016/j.physletb.2005.10.075} {\bibfield
  {journal} {\bibinfo  {journal} {Phys. Lett.}\ }\textbf {\bibinfo {volume}
  {B632}},\ \bibinfo {pages} {543} (\bibinfo {year} {2006})},\ \Eprint
  {http://arxiv.org/abs/hep-th/0502203} {arXiv:hep-th/0502203 [hep-th]}
  \BibitemShut {NoStop}%
\bibitem [{\citenamefont {Buniy}\ \emph {et~al.}(2006)\citenamefont {Buniy},
  \citenamefont {Hsu},\ and\ \citenamefont {Murray}}]{Buniy:2006xf}%
  \BibitemOpen
  \bibfield  {author} {\bibinfo {author} {\bibfnamefont {R.~V.}\ \bibnamefont
  {Buniy}}, \bibinfo {author} {\bibfnamefont {S.~D.~H.}\ \bibnamefont {Hsu}}, \
  and\ \bibinfo {author} {\bibfnamefont {B.~M.}\ \bibnamefont {Murray}},\
  }\href {\doibase 10.1103/PhysRevD.74.063518} {\bibfield  {journal} {\bibinfo
  {journal} {Phys. Rev.}\ }\textbf {\bibinfo {volume} {D74}},\ \bibinfo {pages}
  {063518} (\bibinfo {year} {2006})},\ \Eprint
  {http://arxiv.org/abs/hep-th/0606091} {arXiv:hep-th/0606091 [hep-th]}
  \BibitemShut {NoStop}%
\bibitem [{\citenamefont {Glampedakis}\ and\ \citenamefont
  {Kennefick}(2002)}]{Glampedakis2002}%
  \BibitemOpen
  \bibfield  {author} {\bibinfo {author} {\bibfnamefont {K.}~\bibnamefont
  {Glampedakis}}\ and\ \bibinfo {author} {\bibfnamefont {D.}~\bibnamefont
  {Kennefick}},\ }\href {\doibase 10.1103/physrevd.66.044002} {\bibfield
  {journal} {\bibinfo  {journal} {Physical Review D}\ }\textbf {\bibinfo
  {volume} {66}} (\bibinfo {year} {2002}),\
  10.1103/physrevd.66.044002}\BibitemShut {NoStop}%
\bibitem [{\citenamefont {{Blanchet}}(2014)}]{2014LRR....17....2B}%
  \BibitemOpen
  \bibfield  {author} {\bibinfo {author} {\bibfnamefont {L.}~\bibnamefont
  {{Blanchet}}},\ }\href {\doibase 10.12942/lrr-2014-2} {\bibfield  {journal}
  {\bibinfo  {journal} {Living Reviews in Relativity}\ }\textbf {\bibinfo
  {volume} {17}},\ \bibinfo {eid} {2} (\bibinfo {year} {2014})},\ \Eprint
  {http://arxiv.org/abs/1310.1528} {arXiv:1310.1528 [gr-qc]} \BibitemShut
  {NoStop}%
\bibitem [{\citenamefont {Kozai}(1962)}]{Kozai:1962zz}%
  \BibitemOpen
  \bibfield  {author} {\bibinfo {author} {\bibfnamefont {Y.}~\bibnamefont
  {Kozai}},\ }\href {\doibase 10.1086/108790} {\bibfield  {journal} {\bibinfo
  {journal} {Astron. J.}\ }\textbf {\bibinfo {volume} {67}},\ \bibinfo {pages}
  {591} (\bibinfo {year} {1962})}\BibitemShut {NoStop}%
\bibitem [{\citenamefont {{Lidov}}(1962)}]{1962P&SS....9..719L}%
  \BibitemOpen
  \bibfield  {author} {\bibinfo {author} {\bibfnamefont {M.~L.}\ \bibnamefont
  {{Lidov}}},\ }\href {\doibase 10.1016/0032-0633(62)90129-0} {\bibfield
  {journal} {\bibinfo  {journal} {Planetary and Space Science}\ }\textbf
  {\bibinfo {volume} {9}},\ \bibinfo {pages} {719} (\bibinfo {year}
  {1962})}\BibitemShut {NoStop}%
\bibitem [{\citenamefont {Diemer}\ and\ \citenamefont
  {Smolarek}(2013)}]{DiemerSmolarek:2013mwv}%
  \BibitemOpen
  \bibfield  {author} {\bibinfo {author} {\bibfnamefont {V.}~\bibnamefont
  {Diemer}}\ and\ \bibinfo {author} {\bibfnamefont {E.}~\bibnamefont
  {Smolarek}},\ }\href {\doibase 10.1088/0264-9381/30/17/175014} {\bibfield
  {journal} {\bibinfo  {journal} {Class. Quant. Grav.}\ }\textbf {\bibinfo
  {volume} {30}},\ \bibinfo {pages} {175014} (\bibinfo {year} {2013})},\
  \Eprint {http://arxiv.org/abs/1302.1705} {arXiv:1302.1705 [gr-qc]}
  \BibitemShut {NoStop}%
\bibitem [{\citenamefont {Mikkola}\ and\ \citenamefont
  {Merritt}(2006)}]{MikkolaMerritt2006}%
  \BibitemOpen
  \bibfield  {author} {\bibinfo {author} {\bibfnamefont {S.}~\bibnamefont
  {Mikkola}}\ and\ \bibinfo {author} {\bibfnamefont {D.}~\bibnamefont
  {Merritt}},\ }\href {\doibase 10.1111/j.1365-2966.2006.10854.x} {\bibfield
  {journal} {\bibinfo  {journal} {Monthly Notices of the Royal Astronomical
  Society}\ }\textbf {\bibinfo {volume} {372}},\ \bibinfo {pages} {219}
  (\bibinfo {year} {2006})},\ \Eprint
  {http://arxiv.org/abs/http://oup.prod.sis.lan/mnras/article-pdf/372/1/219/5780989/mnras0372-0219.pdf}
  {http://oup.prod.sis.lan/mnras/article-pdf/372/1/219/5780989/mnras0372-0219.pdf}
  \BibitemShut {NoStop}%
\bibitem [{\citenamefont {Mikkola}\ and\ \citenamefont
  {Merritt}(2008)}]{mikkola2008implementing}%
  \BibitemOpen
  \bibfield  {author} {\bibinfo {author} {\bibfnamefont {S.}~\bibnamefont
  {Mikkola}}\ and\ \bibinfo {author} {\bibfnamefont {D.}~\bibnamefont
  {Merritt}},\ }\href@noop {} {\bibfield  {journal} {\bibinfo  {journal} {The
  Astronomical Journal}\ }\textbf {\bibinfo {volume} {135}},\ \bibinfo {pages}
  {2398} (\bibinfo {year} {2008})}\BibitemShut {NoStop}%
\bibitem [{\citenamefont {Husa}\ \emph {et~al.}(2016)\citenamefont {Husa},
  \citenamefont {Khan}, \citenamefont {Hannam}, \citenamefont {P\"urrer},
  \citenamefont {Ohme}, \citenamefont {Forteza},\ and\ \citenamefont
  {Boh\'e}}]{PhysRevD.93.044006}%
  \BibitemOpen
  \bibfield  {author} {\bibinfo {author} {\bibfnamefont {S.}~\bibnamefont
  {Husa}}, \bibinfo {author} {\bibfnamefont {S.}~\bibnamefont {Khan}}, \bibinfo
  {author} {\bibfnamefont {M.}~\bibnamefont {Hannam}}, \bibinfo {author}
  {\bibfnamefont {M.}~\bibnamefont {P\"urrer}}, \bibinfo {author}
  {\bibfnamefont {F.}~\bibnamefont {Ohme}}, \bibinfo {author} {\bibfnamefont
  {X.~J.}\ \bibnamefont {Forteza}}, \ and\ \bibinfo {author} {\bibfnamefont
  {A.}~\bibnamefont {Boh\'e}},\ }\href {\doibase 10.1103/PhysRevD.93.044006}
  {\bibfield  {journal} {\bibinfo  {journal} {Phys. Rev. D}\ }\textbf {\bibinfo
  {volume} {93}},\ \bibinfo {pages} {044006} (\bibinfo {year}
  {2016})}\BibitemShut {NoStop}%
\bibitem [{\citenamefont {{Jani}}\ \emph {et~al.}(2016)\citenamefont {{Jani}},
  \citenamefont {{Healy}}, \citenamefont {{Clark}}, \citenamefont {{London}},
  \citenamefont {{Laguna}},\ and\ \citenamefont
  {{Shoemaker}}}]{2016CQGra..33t4001J}%
  \BibitemOpen
  \bibfield  {author} {\bibinfo {author} {\bibfnamefont {K.}~\bibnamefont
  {{Jani}}}, \bibinfo {author} {\bibfnamefont {J.}~\bibnamefont {{Healy}}},
  \bibinfo {author} {\bibfnamefont {J.~A.}\ \bibnamefont {{Clark}}}, \bibinfo
  {author} {\bibfnamefont {L.}~\bibnamefont {{London}}}, \bibinfo {author}
  {\bibfnamefont {P.}~\bibnamefont {{Laguna}}}, \ and\ \bibinfo {author}
  {\bibfnamefont {D.}~\bibnamefont {{Shoemaker}}},\ }\href {\doibase
  10.1088/0264-9381/33/20/204001} {\bibfield  {journal} {\bibinfo  {journal}
  {Classical and Quantum Gravity}\ }\textbf {\bibinfo {volume} {33}},\ \bibinfo
  {eid} {204001} (\bibinfo {year} {2016})},\ \Eprint
  {http://arxiv.org/abs/1605.03204} {arXiv:1605.03204 [gr-qc]} \BibitemShut
  {NoStop}%
\bibitem [{\citenamefont {Blanchet}\ \emph {et~al.}(2010)\citenamefont
  {Blanchet}, \citenamefont {Detweiler}, \citenamefont {Le~Tiec},\ and\
  \citenamefont {Whiting}}]{PhysRevD.81.064004}%
  \BibitemOpen
  \bibfield  {author} {\bibinfo {author} {\bibfnamefont {L.}~\bibnamefont
  {Blanchet}}, \bibinfo {author} {\bibfnamefont {S.}~\bibnamefont {Detweiler}},
  \bibinfo {author} {\bibfnamefont {A.}~\bibnamefont {Le~Tiec}}, \ and\
  \bibinfo {author} {\bibfnamefont {B.~F.}\ \bibnamefont {Whiting}},\ }\href
  {\doibase 10.1103/PhysRevD.81.064004} {\bibfield  {journal} {\bibinfo
  {journal} {Phys. Rev. D}\ }\textbf {\bibinfo {volume} {81}},\ \bibinfo
  {pages} {064004} (\bibinfo {year} {2010})}\BibitemShut {NoStop}%
\bibitem [{\citenamefont {Maggiore}(2008)}]{maggiore2008gravitational}%
  \BibitemOpen
  \bibfield  {author} {\bibinfo {author} {\bibfnamefont {M.}~\bibnamefont
  {Maggiore}},\ }\href {https://books.google.com/books?id=mk-1DAAAQBAJ} {\emph
  {\bibinfo {title} {Gravitational Waves: Volume 1: Theory and Experiments}}},\
  Gravitational Waves\ (\bibinfo  {publisher} {OUP Oxford},\ \bibinfo {year}
  {2008})\BibitemShut {NoStop}%
\bibitem [{\citenamefont {GWIC}(2019)}]{2019:3GWhitePaper}%
  \BibitemOpen
  \bibfield  {author} {\bibinfo {author} {\bibnamefont {GWIC}},\ }\href
  {https://gwic.ligo.org/3Gsubcomm/documents/3G-observatory-science-case.pdf}
  {\enquote {\bibinfo {title} {Gravitational-wave astronomy with the
  next-generation earth-based observatories},}\ } (\bibinfo {year}
  {2019})\BibitemShut {NoStop}%
\bibitem [{\citenamefont {{Reitze}}\ \emph {et~al.}(2019)\citenamefont
  {{Reitze}}, \citenamefont {{Adhikari}}, \citenamefont {{Ballmer}},
  \citenamefont {{Barish}}, \citenamefont {{Barsotti}}, \citenamefont
  {{Billingsley}}, \citenamefont {{Brown}}, \citenamefont {{Chen}},
  \citenamefont {{Coyne}}, \citenamefont {{Eisenstein}}, \citenamefont
  {{Evans}}, \citenamefont {{Fritschel}}, \citenamefont {{Hall}}, \citenamefont
  {{Lazzarini}}, \citenamefont {{Lovelace}}, \citenamefont {{Read}},
  \citenamefont {{Sathyaprakash}}, \citenamefont {{Shoemaker}}, \citenamefont
  {{Smith}}, \citenamefont {{Torrie}}, \citenamefont {{Vitale}}, \citenamefont
  {{Weiss}}, \citenamefont {{Wipf}},\ and\ \citenamefont
  {{Zucker}}}]{2019BAAS...51g..35R}%
  \BibitemOpen
  \bibfield  {author} {\bibinfo {author} {\bibfnamefont {D.}~\bibnamefont
  {{Reitze}}}, \bibinfo {author} {\bibfnamefont {R.~X.}\ \bibnamefont
  {{Adhikari}}}, \bibinfo {author} {\bibfnamefont {S.}~\bibnamefont
  {{Ballmer}}}, \bibinfo {author} {\bibfnamefont {B.}~\bibnamefont {{Barish}}},
  \bibinfo {author} {\bibfnamefont {L.}~\bibnamefont {{Barsotti}}}, \bibinfo
  {author} {\bibfnamefont {G.}~\bibnamefont {{Billingsley}}}, \bibinfo {author}
  {\bibfnamefont {D.~A.}\ \bibnamefont {{Brown}}}, \bibinfo {author}
  {\bibfnamefont {Y.}~\bibnamefont {{Chen}}}, \bibinfo {author} {\bibfnamefont
  {D.}~\bibnamefont {{Coyne}}}, \bibinfo {author} {\bibfnamefont
  {R.}~\bibnamefont {{Eisenstein}}}, \bibinfo {author} {\bibfnamefont
  {M.}~\bibnamefont {{Evans}}}, \bibinfo {author} {\bibfnamefont
  {P.}~\bibnamefont {{Fritschel}}}, \bibinfo {author} {\bibfnamefont {E.~D.}\
  \bibnamefont {{Hall}}}, \bibinfo {author} {\bibfnamefont {A.}~\bibnamefont
  {{Lazzarini}}}, \bibinfo {author} {\bibfnamefont {G.}~\bibnamefont
  {{Lovelace}}}, \bibinfo {author} {\bibfnamefont {J.}~\bibnamefont {{Read}}},
  \bibinfo {author} {\bibfnamefont {B.~S.}\ \bibnamefont {{Sathyaprakash}}},
  \bibinfo {author} {\bibfnamefont {D.}~\bibnamefont {{Shoemaker}}}, \bibinfo
  {author} {\bibfnamefont {J.}~\bibnamefont {{Smith}}}, \bibinfo {author}
  {\bibfnamefont {C.}~\bibnamefont {{Torrie}}}, \bibinfo {author}
  {\bibfnamefont {S.}~\bibnamefont {{Vitale}}}, \bibinfo {author}
  {\bibfnamefont {R.}~\bibnamefont {{Weiss}}}, \bibinfo {author} {\bibfnamefont
  {C.}~\bibnamefont {{Wipf}}}, \ and\ \bibinfo {author} {\bibfnamefont
  {M.}~\bibnamefont {{Zucker}}},\ }in\ \href@noop {} {\emph {\bibinfo
  {booktitle} {\baas}}},\ Vol.~\bibinfo {volume} {51}\ (\bibinfo {year}
  {2019})\ p.~\bibinfo {pages} {35},\ \Eprint {http://arxiv.org/abs/1907.04833}
  {arXiv:1907.04833 [astro-ph.IM]} \BibitemShut {NoStop}%
\bibitem [{Cos()}]{CosmicExploreWebsite}%
  \BibitemOpen
  \href@noop {} {\enquote {\bibinfo {title} {Cosmic explorer website},}\
  }\bibinfo {howpublished} {\url{https://cosmicexplorer.org}},\ \bibinfo {note}
  {accessed: 2020-07-09}\BibitemShut {NoStop}%
\bibitem [{\citenamefont {Yu}\ \emph {et~al.}(2018)\citenamefont {Yu} \emph
  {et~al.}}]{Yu:2017zgi}%
  \BibitemOpen
  \bibfield  {author} {\bibinfo {author} {\bibfnamefont {H.}~\bibnamefont {Yu}}
  \emph {et~al.},\ }\href {\doibase 10.1103/PhysRevLett.120.141102} {\bibfield
  {journal} {\bibinfo  {journal} {Phys. Rev. Lett.}\ }\textbf {\bibinfo
  {volume} {120}},\ \bibinfo {pages} {141102} (\bibinfo {year} {2018})},\
  \Eprint {http://arxiv.org/abs/1712.05417} {arXiv:1712.05417 [astro-ph.IM]}
  \BibitemShut {NoStop}%
\bibitem [{\citenamefont {{Colpi}}\ \emph {et~al.}(2019)\citenamefont
  {{Colpi}}, \citenamefont {{Holley-Bockelmann}}, \citenamefont {{Bogdanovic}},
  \citenamefont {{Natarajan}}, \citenamefont {{Bellovary}}, \citenamefont
  {{Sesana}}, \citenamefont {{Tremmel}}, \citenamefont {{Schnittman}},
  \citenamefont {{Comerford}}, \citenamefont {{Barausse}}, \citenamefont
  {{Berti}}, \citenamefont {{Volonteri}}, \citenamefont {{Khan}}, \citenamefont
  {{McWilliams}}, \citenamefont {{Burke-Spolaor}}, \citenamefont {{Hazboun}},
  \citenamefont {{Conklin}}, \citenamefont {{Mueller}},\ and\ \citenamefont
  {{Larson}}}]{ColpiWP}%
  \BibitemOpen
  \bibfield  {author} {\bibinfo {author} {\bibfnamefont {M.}~\bibnamefont
  {{Colpi}}}, \bibinfo {author} {\bibfnamefont {K.}~\bibnamefont
  {{Holley-Bockelmann}}}, \bibinfo {author} {\bibfnamefont {T.}~\bibnamefont
  {{Bogdanovic}}}, \bibinfo {author} {\bibfnamefont {P.}~\bibnamefont
  {{Natarajan}}}, \bibinfo {author} {\bibfnamefont {J.}~\bibnamefont
  {{Bellovary}}}, \bibinfo {author} {\bibfnamefont {A.}~\bibnamefont
  {{Sesana}}}, \bibinfo {author} {\bibfnamefont {M.}~\bibnamefont {{Tremmel}}},
  \bibinfo {author} {\bibfnamefont {J.}~\bibnamefont {{Schnittman}}}, \bibinfo
  {author} {\bibfnamefont {J.}~\bibnamefont {{Comerford}}}, \bibinfo {author}
  {\bibfnamefont {E.}~\bibnamefont {{Barausse}}}, \bibinfo {author}
  {\bibfnamefont {E.}~\bibnamefont {{Berti}}}, \bibinfo {author} {\bibfnamefont
  {M.}~\bibnamefont {{Volonteri}}}, \bibinfo {author} {\bibfnamefont
  {F.}~\bibnamefont {{Khan}}}, \bibinfo {author} {\bibfnamefont
  {S.}~\bibnamefont {{McWilliams}}}, \bibinfo {author} {\bibfnamefont
  {S.}~\bibnamefont {{Burke-Spolaor}}}, \bibinfo {author} {\bibfnamefont
  {J.}~\bibnamefont {{Hazboun}}}, \bibinfo {author} {\bibfnamefont
  {J.}~\bibnamefont {{Conklin}}}, \bibinfo {author} {\bibfnamefont
  {G.}~\bibnamefont {{Mueller}}}, \ and\ \bibinfo {author} {\bibfnamefont
  {S.}~\bibnamefont {{Larson}}},\ }\href@noop {} {\bibfield  {journal}
  {\bibinfo  {journal} {arXiv e-prints}\ ,\ \bibinfo {eid} {arXiv:1903.06867}}
  (\bibinfo {year} {2019})},\ \Eprint {http://arxiv.org/abs/1903.06867}
  {arXiv:1903.06867 [astro-ph.GA]} \BibitemShut {NoStop}%
\bibitem [{\citenamefont {Torres}\ \emph {et~al.}(1998)\citenamefont {Torres},
  \citenamefont {Romero},\ and\ \citenamefont {Anchordoqui}}]{Torres:1998xd}%
  \BibitemOpen
  \bibfield  {author} {\bibinfo {author} {\bibfnamefont {D.~F.}\ \bibnamefont
  {Torres}}, \bibinfo {author} {\bibfnamefont {G.~E.}\ \bibnamefont {Romero}},
  \ and\ \bibinfo {author} {\bibfnamefont {L.~A.}\ \bibnamefont
  {Anchordoqui}},\ }\href {\doibase 10.1103/PhysRevD.58.123001} {\bibfield
  {journal} {\bibinfo  {journal} {Phys. Rev.}\ }\textbf {\bibinfo {volume}
  {D58}},\ \bibinfo {pages} {123001} (\bibinfo {year} {1998})},\ \Eprint
  {http://arxiv.org/abs/astro-ph/9802106} {arXiv:astro-ph/9802106 [astro-ph]}
  \BibitemShut {NoStop}%
\bibitem [{\citenamefont {Graham}\ \emph {et~al.}(2020)\citenamefont {Graham}
  \emph {et~al.}}]{Graham:2020gwr}%
  \BibitemOpen
  \bibfield  {author} {\bibinfo {author} {\bibfnamefont {M.~J.}\ \bibnamefont
  {Graham}} \emph {et~al.},\ }\href {\doibase 10.1103/PhysRevLett.124.251102}
  {\bibfield  {journal} {\bibinfo  {journal} {Phys. Rev. Lett.}\ }\textbf
  {\bibinfo {volume} {124}},\ \bibinfo {pages} {251102} (\bibinfo {year}
  {2020})},\ \Eprint {http://arxiv.org/abs/2006.14122} {arXiv:2006.14122
  [astro-ph.HE]} \BibitemShut {NoStop}%
\bibitem [{\citenamefont {Dai}\ and\ \citenamefont
  {Stojkovic}(2019)}]{Dai:2019mse}%
  \BibitemOpen
  \bibfield  {author} {\bibinfo {author} {\bibfnamefont {D.-C.}\ \bibnamefont
  {Dai}}\ and\ \bibinfo {author} {\bibfnamefont {D.}~\bibnamefont
  {Stojkovic}},\ }\href {\doibase 10.1103/PhysRevD.100.083513} {\bibfield
  {journal} {\bibinfo  {journal} {Phys. Rev. D}\ }\textbf {\bibinfo {volume}
  {100}},\ \bibinfo {pages} {083513} (\bibinfo {year} {2019})},\ \Eprint
  {http://arxiv.org/abs/1910.00429} {arXiv:1910.00429 [gr-qc]} \BibitemShut
  {NoStop}%
\bibitem [{\citenamefont {Hochberg}\ and\ \citenamefont
  {Kephart}(1991)}]{Hochberg:1991tz}%
  \BibitemOpen
  \bibfield  {author} {\bibinfo {author} {\bibfnamefont {D.}~\bibnamefont
  {Hochberg}}\ and\ \bibinfo {author} {\bibfnamefont {T.~W.}\ \bibnamefont
  {Kephart}},\ }\href {\doibase 10.1016/0370-2693(91)91593-K} {\bibfield
  {journal} {\bibinfo  {journal} {Phys. Lett.}\ }\textbf {\bibinfo {volume}
  {B268}},\ \bibinfo {pages} {377} (\bibinfo {year} {1991})}\BibitemShut
  {NoStop}%
\bibitem [{\citenamefont {Duplessis}\ and\ \citenamefont
  {Easson}(2015)}]{Duplessis:2015xva}%
  \BibitemOpen
  \bibfield  {author} {\bibinfo {author} {\bibfnamefont {F.}~\bibnamefont
  {Duplessis}}\ and\ \bibinfo {author} {\bibfnamefont {D.~A.}\ \bibnamefont
  {Easson}},\ }\href {\doibase 10.1103/PhysRevD.92.043516} {\bibfield
  {journal} {\bibinfo  {journal} {Phys. Rev.}\ }\textbf {\bibinfo {volume}
  {D92}},\ \bibinfo {pages} {043516} (\bibinfo {year} {2015})},\ \Eprint
  {http://arxiv.org/abs/1506.00988} {arXiv:1506.00988 [gr-qc]} \BibitemShut
  {NoStop}%
\bibitem [{\citenamefont {Dent}\ \emph {et~al.}(2017)\citenamefont {Dent},
  \citenamefont {Easson}, \citenamefont {Kephart},\ and\ \citenamefont
  {White}}]{Dent:2016efw}%
  \BibitemOpen
  \bibfield  {author} {\bibinfo {author} {\bibfnamefont {J.~B.}\ \bibnamefont
  {Dent}}, \bibinfo {author} {\bibfnamefont {D.~A.}\ \bibnamefont {Easson}},
  \bibinfo {author} {\bibfnamefont {T.~W.}\ \bibnamefont {Kephart}}, \ and\
  \bibinfo {author} {\bibfnamefont {S.~C.}\ \bibnamefont {White}},\ }\href
  {\doibase 10.1142/S0218271817501176} {\bibfield  {journal} {\bibinfo
  {journal} {Int. J. Mod. Phys.}\ }\textbf {\bibinfo {volume} {D26}},\ \bibinfo
  {pages} {1750117} (\bibinfo {year} {2017})},\ \Eprint
  {http://arxiv.org/abs/1608.00589} {arXiv:1608.00589 [gr-qc]} \BibitemShut
  {NoStop}%
\bibitem [{\citenamefont {Hochberg}\ and\ \citenamefont
  {Kephart}(1993)}]{Hochberg:1992du}%
  \BibitemOpen
  \bibfield  {author} {\bibinfo {author} {\bibfnamefont {D.}~\bibnamefont
  {Hochberg}}\ and\ \bibinfo {author} {\bibfnamefont {T.~W.}\ \bibnamefont
  {Kephart}},\ }\href {\doibase 10.1103/PhysRevLett.70.2665} {\bibfield
  {journal} {\bibinfo  {journal} {Phys. Rev. Lett.}\ }\textbf {\bibinfo
  {volume} {70}},\ \bibinfo {pages} {2665} (\bibinfo {year} {1993})},\ \Eprint
  {http://arxiv.org/abs/gr-qc/9211006} {arXiv:gr-qc/9211006 [gr-qc]}
  \BibitemShut {NoStop}%
\end{thebibliography}%


\end{document}